\documentclass[12pt]{article}

\usepackage{amsmath}
\usepackage{amssymb}

\usepackage{graphicx}

\usepackage{cite}

\makeatletter
\renewcommand{\@biblabel}[1]{\quad#1.}
\makeatother

\newcommand{\be}{\begin{equation}}
\newcommand{\ee}{\end{equation}}
\newcommand{\Dlt}{\Delta}
\newcommand{\dlt}{\delta}

\newcommand{\vp}{\varphi}

\newcommand{\ra}{\rightarrow}
\newcommand{\sgm}{\sigma}

\begin{document}

\begin{center}

{\Large{\bf 
Utility Rate Equations of Group Population Dynamics in Biological
and Social Systems} \\ [5mm]

V.I. Yukalov$^{1,2*}$, E.P. Yukalova$^{1,3}$, and
D. Sornette$^{1,4}$} \\ [3mm]

{\it
$^1$Department of Management, Technology and Economics, \\
ETH Z\"urich, Swiss Federal Institute of Technology, \\
Z\"urich CH-8092, Switzerland \\ [3mm]

$^2$Bogolubov Laboratory of Theoretical Physics, \\
Joint Institute for Nuclear Research, Dubna 141980, Russia \\ [3mm]

$^3$Laboratory of Information Technologies, \\
Joint Institute for Nuclear Research, Dubna 141980, Russia \\ [3mm]

$^4$Swiss Finance Institute, c/o University of Geneva, \\
40 blvd. Du Pont d'Arve, CH 1211 Geneva 4, Switzerland} \\ [3mm]

\end{center}

\vskip 3cm

\begin{abstract}
We present a novel system of equations to describe the evolution of self-organized
structured societies (biological or human) composed of several trait groups. The
suggested approach is based on the combination of ideas employed in the theory
of biological populations, system theory, and utility theory. The evolution equations
are defined as utility rate equations, whose parameters are characterized by the
utility of each group with respect to the society as a whole and by the mutual
utilities of groups with respect to each other. We analyze in detail the cases of
two groups (cooperators and defectors) and of three groups (cooperators, defectors,
and regulators) and find that, in a self-organized society, neither defectors nor
regulators can overpass the maximal fractions of about $10\%$ each. This is in
agreement with the data for bee and ant colonies. The classification of societies
by their distance from equilibrium is proposed. We apply the formalism to rank the
countries according to the introduced metric quantifying their relative stability,
which depends on the cost of defectors and regulators as well as their respective
population fractions. We find a remarkable concordance with more standard economic
ranking based, for instance, on GDP per capita.
\end{abstract}

\vskip 2cm
$^*${\bf E-mail}: yukalov@theor.jinr.ru

\newpage

\section*{Introduction}

Biological and human social systems have many common features allowing
considering them on common grounds \cite{1}. Actually, biological systems are
nothing but particular kinds of social systems. And vice versa, human societies
are particular types of biological systems. Therefore, the term social system
can be applied to both examples of societies.

By definition, a biological or social system is an ensemble of entities
characterized by their structure controlling enduring and relatively stable
patterns of relationship between different groups of the society \cite{2,3,4,5,6}.
A social system is structured into several groups defined through their actions
in mutual relations. The characteristics of these relations can change in the
long run but, over limited time spans, they can be treated as fixed.

Any social system exists in an environment that certainly
influences its properties. However, the definition of a social
system presupposes that the influence of the environment does not
preclude to identify the system as an entity, with its properties
that are more or less specified for a sufficiently long time
during which the system is well defined as such. That is, the
notion of a social system implies that it can be defined as
a self-organized entity, with given characteristics, which lives a
sufficiently long time in order to be specified as a particular
system. The system lifetime is much longer that the typical
interaction time of its constituents, though it can be limited from
above by the time scales at which the systems characteristics
essentially vary. In that sense, the system lifetime is
intermediate, being between the local variation time of its
components and the time of the global change of its properties.

The structure and behavior of different societies, though
being specific to each of them, nevertheless do possess general
features that can be classified using mathematical methods,
such that the variety of behaviors can be mapped onto differences
in the system parameters and initial conditions. The concept, that
social systems exhibit general reproducible behaviors and
properties, has been emphasized in system theory  \cite{1,7,8,9,10,11,12,13},
which abstracts and considers a system as a connected set of
interacting parts. The main goal of system theory is to study
general principles of how systems function that can be applied
to all types of systems from a given class.

Bertalanffy \cite{1}, who coined the term {\it system theory},
emphasized that biological systems have much in common with human
social systems, as well as with separate biological organisms that
can be treated as biological societies of different organs and
cells. Dynamical models of biological organisms are also senseful
at intermediate time scales that are longer than the characteristic
functional times of the considered organism constituents, but
shorter than the whole organism lifetimes \cite{14,15}. This is because
during the whole organism lifetime, the parameters of the system can
essentially vary, reflecting the aging processes. In the same way, the
dynamical models of social systems are appropriate for describing
intermediate time scales, during which the system parameters do
not essentially change. Generally, the model structure can even
be preserved for rather long times, of the order of the society
lifetime, provided that the society parameters are slowly varying,
so that it is possible to separate the system dynamics at different
time scales, resorting to the scale separation approach \cite{16,17}.

The classification, analysis, understanding and prediction
of the  evolution of biological societies, including human societies,
is an old problem that has been treated in a voluminous literature
starting from Darwin \cite{18,19}. Societies are usually
structured into groups representing particular traits or strategies,
so that such groups are termed {\it trait groups}, or
{\it strategy groups}. Each group consists of agents with a specific
trait and/or strategy. Typical group representatives are {\it collaborators}
and {\it defectors}. The former collaborate with each other,
contributing to the benefits of the whole society, motivated by
self-interest and/or altruistic behavior. The latter, sometimes
called free riders, do not contribute to the society but,
on the contrary, exploit it, benefiting from it without enduring costs
or making sacrifices.

Biological explanations of cooperation are based on kin altruism,
reciprocal altruism, reputation or signaling, and mutualism, all of which
apply to human as well as nonhuman species \cite{20}. In humans, altruism
can be enforced by internal norms  \cite{21,22,23,24}. According to the
standard description, in a society, consisting of just two groups, cooperators
and defectors, the latter always outcompete altruists \cite{25}. The
cooperation can be stabilized by the punishment of defectors \cite{26,27,28}.
There are, however, examples  \cite{29} when punishment decreases cooperation.
While costly punishment, termed altruistic, may facilitate cooperation, at
the same time, it can be interpreted as not so altruistic but rather selfish,
because the original motivation behind punishment could be to retrieve
deserved payoffs from their own contributions on the long run, which is
a selfish incentive \cite{30,31,32}. Punishment can also act indirectly,
since the defectors, outcompeting cooperators inside a society, weaken the
competitive ability of the society as a whole, as is considered in
multilevel selection theory  \cite{33,34,35,25,36}.

The punishment of defectors can be realized directly by other members of
the society. Since the ability of punishing can also be treated as a
specific trait or strategy, it can be associated with a specific
group of {\it punishers}. This is especially evident in
complex societies, where punishers really form separate groups.
For instance, in human societies, punishers are represented by
police, law-enforcing services, and the army, which form
what can be called the punisher group. The consideration
of such trait groups does not forbid for the manifolds of cooperators
and punishers to intersect \cite{37,38}.

Sometimes, one considers one more specific trait group whose members are
alienated from the society. The agents of such alienated groups, depending
on their particular features, are termed by different names, as strangers
\cite{39}, loners \cite{40}, outsiders \cite{38}, and by other similar terms.
These {\it outsiders} do not contribute to the society, but exploit
cooperators and are prosecuted by punishers. Their difference from defectors
is that they enter the society from outside. There are plenty of examples
of such outsiders in biological systems, where they are associated with
pathogens or parasites  \cite{41}.

The evolution of groups is usually studied within evolutionary game
theory \cite{42,43,44,45}, whose continuous representation is given
by the replicator equation \cite{46,47,48}
\be
\label{1}
\frac{dx_i}{dt} = ( f_i - \overline f ) x_i \;  .
\ee
Here $x_i$ is the population fraction of group $i$ and $f_i$ is the group
fitness, and ${\overline f}  \equiv \sum_i f_i x_i$ is the average fitness
estimated over all groups. The group fractions are defined on the simplex.
The trait-group fitness is often represented by the quadratic Lande-Arnold
form widely used in describing the evolution of biological species, with
the coefficients defined by calibration \cite{49,50,51}.
The most interesting information that can be extracted from this dynamics
is the existence and properties of evolutionary stable strategies related
to the stable stationary states of the dynamical system.

As is evident from the structure of the replicator equation (\ref{1}), fitness
is a quantity inversely proportional to time. This is concretized in the
definition of fitness as being the product of viability and fecundity rate
\cite{52,53,54,55}. However, for short, one uses the term fitness, but not
fitness rate. Similarly, below we shall use the term utility, but not utility
rate. This will not be important for the general consideration, where we shall
always employ reduced dimensionless quantities. But we shall keep this in mind
in those concrete cases, where numerical values of parameters are important.

As said above, if the society consists of two groups only, cooperators
$(x_1)$ and defectors $(x_2)$, the latter always outperform the former,
so that the sole evolutionary stable state is $x^*_1 = 0, x^*_2 = 1$.
This unrealistic conclusion is caused by a tacit assumption that there
exist unlimited resources supplied from somewhere outside.

Contrary to this, we consider a {\it self-organized} society, whose means
of survival are those that are produced inside the society itself.
A closed self-organized society cannot exist by just being composed of
defectors because they will have no means for subsistence. To correct
such an unrealistic conclusion, one could introduce punishers. But, from
a self-organization view point, the introduction of punishers is not
compulsory since, even without them, defectors cannot proliferate without
bounds because of the lack of resources for their existence \cite{56,57}.

It is necessary to note that the non-survival of cooperators can disappear,
when one considers interactions on a network. Thus, computer simulations
of cooperator games on lattices have demonstrated that cooperation can evolve
in structural populations when cooperators interact more frequently with each
other than with defectors, and share benefits of mutual cooperation
\cite{Rap,Tri}, or when individuals interact at different rates with each other
by controlling the frequency distribution of the number of their partners
\cite{Wu,Cao}. Cooperators survive on a network in the presence of modularity
describing the separation of the population into clusters of different traits
\cite{Mar}. High modularity indicates a network with a strong clustered structure,
where the interactions of individuals belonging to different clusters do not
occur \cite{New}. Modularity also arises, when the rate of interaction between
different groups is not the same for all groups \cite{New,Gir}. Modularity
favours cooperation by limiting the interactions of individuals to the members
of the same community \cite{Mar}.

The principal difference between our article and these previous computer
simulations is that we demonstrate analytically and parsimoniously the survival
of cooperation without considering numerical calculations on networks. We show
that the finiteness of resources is sufficient to limit the proliferation of
defectors, without the need for heterogeneous interactions and clustering. While
the latter ingredients are certainly relevant in many situations, our focus on the
finiteness of resources is justified because it is often the first and dominant
variable influencing population dynamics.

In the present paper, we combine some of the ideas used in the theory of
biological populations, system theory, and utility theory and suggest a novel
model of group evolution in a complex {\it self-organized} society, which strives
to be more realistic than those usually studied in game theory. The main idea of
our approach is in taking into account the bounded rationality of realistic
agents \cite{58,59}. Accordingly, the utility of an agent is not an absolutely
fixed notion, but is instead a relative characteristic defined with respect to
the existing constraints. In our approach, instead of just one utility function
for each group, we introduce relative utilities of a group with respect to the
society as a whole as well as mutual utilities with respect to each other.

As an example, we consider a complex structured society composed of the trait
groups representing four types of typical agents: {\it cooperators, defectors,
regulators}, and {\it outsiders}. Concrete numerical analysis is given for the
societies composed of two groups ({\it cooperators and defectors}) and of three
groups ({\it cooperators, defectors}, and {\it regulators}). The role of outsiders
will be studied in a separate publication.

\section*{Evolution equations}

\subsection*{Holistic system-theory approach}

As we have emphasized in the Introduction, we follow the ideas
underpinning the theory of systems
\cite{1,7,8,9,10,11,12,13}, whose basic point is the holistic approach
for describing systems.  Accordingly, the primary unit is the system as
a whole, as opposed to the reductionist approach, where a system is treated
just as a collection of separate parts. In the holistic approach, a system is
considered as a self-regulating organism, whose homeostasis is regulated by
the system itself tending to reach and maintain a stable state. The regulation
and control are achieved through positive and negative feedbacks. This approach
is rather general, being applicable to different systems, physical, chemical,
engineering, cybernetic, economic, social, and biological. In the present
subsection, we delineate the main idea of the approach, leaving specifications
to the following subsections.

Suppose that we aim at describing the behavior of a system consisting of several
parts that, e.g., are characterized by fractions $x_i$, with $i = 1,2,\ldots$.
The fractions are the reduced quantities
\be
\label{2}
x_i \equiv \frac{N_i}{N_0} \; ,
\ee
whose specifications will be given below, and meanwhile their nature is not
important, since the approach can be applied to various particular cases, as
is mentioned above.

In the reductionsit approach, one characterises each fraction by its parameters
and describes the system as a collection of equations containing only these
parameters specific for separate fractions. An example of such an approach
would apply to the collection of species, each characterised by its fitness and
described by a replicator equation of type (1).

In the holistic system-theory approach, one starts with defining a
{\it system functional} that characterises the system as a whole. Actually,
this approach is standard and is known to be the most general and accurate in
natural sciences, such as physics or chemistry, where a composite system is
characterized by its energy, free energy, or Hamiltonian defining the system
as a whole. In applied sciences involving control theory, a composite system
is characterized by a cost functional. In economic, social, and financial
sciences, it is more customary to characterize a system by the related utility.
All these characteristics, whether {\it energy} $E$, {\it cost functional} $C$,
or {\it utility} $U$ are functionals of the system parts, such as the fractions
$x_i$ depending on time, $x_i = x_i(t)$. These characteristics are particular
examples of the system functional. Generally, all of them are proportional
to each other, thus, $E \sim C \sim -U$. Therefore, it would be possible to
deal with any of those system functionals. In what follows, keeping in mind
biological and social systems, we prefer to work with the {\it system utility}
that is a functional $U = U(\{x_i\})$. Since the variables $x_i(t)$ depend
on time, this can be called a {\it dynamic utility functional} \cite{Weidlich_1}.

Equations, describing the behavior of the system parts, in the holistic system-theory
approach, are derived by varying the system functional, such as energy
or cost functional. Recall again that this is the standard way in natural
sciences, where the evolution equations for system parts are obtained through
the variation of a system functional, e.g., energy, Hamiltonian, or Lagrangian.
Even the Heisenberg commutator equations can be shown \cite{PLA} to be equivalent
to variational equations for the total system Hamiltonian.

The system tends to its stable state that requires the increase of its utility.
Hence, a fraction $x_i$ increases with time if its variation leads to the
increasing system utility. This implies that $x_i$ rises when the variational
derivative $\delta U/ \delta x_i$ is positive, while $x_i$ diminishes with time
if $\delta U/ \delta x_i$ is negative. It is straightforward that this relation,
taking into account possible external influxes $\Phi_i$, can be represented
in the form
\be
\label{3}
 \frac{dx_i}{dt} = \frac{\dlt U}{\dlt x_i} \; x_i + \Phi_i \;  .
\ee

A self-regulating system self-organizes through its internal feedbacks, acting
so that its utility would not diminish. Such feedbacks are defined by the linear
responses
$$
 A_i \equiv \lim_{ \{ x_i\ra 0\} } \; \frac{\dlt U}{\dlt x_i} \;  ,
$$
describing the direct influence of the fraction variations on the system utility.
From this definition, it is clear that, when the increase of the fraction $x_i$
directly results in the increase of the system utility then, the feedback
response $A_i$ is positive. And if the rising $x_i$ leads to the utility
decrease, then the response $A_i$ is negative. In addition to the direct
influence on the system utility of the fraction variations, these fractions
are correlated with each other through the correlation matrix, which is proportional to
$$
   A_{ij} \equiv
\lim_{ \{ x_i\ra 0\} } \; \frac{\dlt^2 U}{\dlt x_i \dlt x_j} ~.
$$
By construction, it is symmetric such that $A_{ij} = A_{ji}$.

From the above discussion, it immediately follows that the general form of the
system utility reads as
\be
\label{4}
  U = \sum_i A_i x_i + \frac{1}{2} \; \sum_{ij} A_{ij} x_i x_j \; .
\ee
Again we recall that such a quadratic form is typical for the system energy
in natural sciences or for the cost functional in control theory and engineering
applications. It is also worth mentioning the Lande-Arnold \cite{49,50,51}
quadratic fitness for biological systems. With the system utility (4), we come
to the evolution equations for the fractions:
\be
\label{5}
\frac{d x_i}{dt} = A_i x_i +  \sum_{j} A_{ij} x_i x_j + \Phi_i \; .
\ee

We stress that the signs of all parameters are uniquely defined by the conditions
of whether the fraction variations increase or decrease the system utility. Thus,
the response $A_i$ is positive if the increase of the fraction $x_i$ increases
the system utility, while $A_i$ is negative when the fraction $x_i$ decreases
this utility.

In this way, the parameter signs are not artificially chosen but are prescribed
by the used holistic approach. Moreover, it would not be correct to say that the
behavior of solutions is predetermined. Since, for instance, though the feedback
of defectors on the society utility is negative ($A_2 < 0$), but the input of
cooperators to the utility of defectors is positive ($A_{21} > 0$). Therefore the
actual behavior of the solution describing defectors is not at all predetermined,
but depends on the delicate balance between the values of all system parameters.

\subsection*{Utility rate equations}

Let us now make concrete the general approach, which was described in the previous
subsection, to the case of social and biological systems. We consider a society
composed of the trait groups, or strategy groups, enumerated by an index
$i = 1,2, \ldots$, each consisting of $N_i$ members. Each group will be
characterized by the utility of the group with respect to the whole society
and with respect to other groups. The concrete form of this dependence will
be defined below.

In general, there can exist an influx $\Phi_i$ from outside to each group
population. Such an influx can represent, for instance, pathogens or viruses
infecting a biological organism. In the case of social systems, the influx
can correspond to foreign invaders and terrorists.

The number of agents in a group can be very large, so that it is
standard to consider the relative fractions $x_i$ of the group
populations normalized to a constant number $N_0$, as in Eq. (2). If the
total population of all groups would be fixed, it would be possible
to interpret the normalization constant as the total population. However,
in general, the group populations can grow or decay so that the total
population is not necessarily fixed. In addition, there can exist external
influxes of populations, not allowing for the conservation of the total population.
Hence, fixing the total population would severely limit the admissible scenarios.
However, it is always possible to choose for the normalization the amount
of the total population at a given time. The most natural way is to accept
for the normalization the total population $N_0 = \sum_i N_i(0)$ at the
initial time $t = 0$, which therefore sets the scale for the population size.

The total population of any society is certainly finite. In that sense, the
variable $x_i$ could be treated as discrete. However, when one considers large
societies, as we keep in mind here, when the population can be of order of
many thousands or millions, it is standard to treat $x_i$ as a continuous
variable.

In the biological literature, one usually characterizes the species by their
fitness. And in the literature on social systems, one often employs the notion
of utility. While the classical economic approach often defines the utility
of an agent as the present value of future consumption, inter-generational
extensions include bequest and the consumption of future generations.
It is often argued that one of the most important measure of success is the
passing of genes to future generation \cite{Dawkins76,FavreSor12}, so that
utility, as a measure of happiness, and fitness, as a measure of success,
become entangled concepts. We prefer to use the term utility, keeping in mind
applications to both biological as well as social systems.

The evolution equations (\ref{5}) possess the form of the Lotka-Volterra
rate equation. However, the basic novelty of our way is that we employ
the system-theory holistic approach, as described above, which suggests a
different interpretation of the coefficients that are defined not by some
interactions between the group members, but by the utility of each group
with respect to other groups and to the whole society. This prescription is
important for correctly identifying the signs of the coefficients. Thus, the
term $A_i$ represents the utility of the $i$-group to the society as a whole
and the terms $A_{ij}$ characterize the mutual utility that the $j$-th group
provides to the $i$-th group.

The evolution equations, we have just formulated, are the
{\it utility rate equations}. For these equations, the number and signs of
the coefficients are uniquely defined by the mutual utility of the groups,
as is explained above.

It is worth emphasizing that the definition of the response $A_i$ as being
not purely intrinsic, but dependent on the relation of the $i$-th population
to the overall system, playing the role of environment for that population,
is not merely admissible, but rather necessary.

The form of Eq. (\ref{5}) is typical of many equations employed for describing
the evolution of different biological species, where the role of the rate $A_i$
is played by the species fitness. As examples, we can mention the replicator
equation (\ref{1}), the Eigen \cite{Eigen,Schuster} and Crow-Kimura
\cite{Crow,Kimura} equations, and other models reviewed in Refs.
\cite{Akin,Evens,Hofbauer,Nowak,Jain}.

The population variation rate, as well as fitness, are never completely
intrinsic, but strongly depend on environment carrying capacity, that is, on
the available resources, their quantity and quality
\cite{Mac,Baranyi,Hatfield,Morris}, the existence of other coexisting species
\cite{Kremer}, and even on the predictability of environmental changes
\cite{Wilbur,Brockelman}. The environment essentially influences the species
fitness that can increase or decrease and sometimes unfavorable environments
result in the fitness drop and species extinction \cite{Lande}.

Recall that the general definition of fitness, from the very beginning
\cite{Haldane,Dobzhansky,Orr}, has always been stressing its dependence on
surrounding: "Fitness is the ability of an organism to survive and reproduce
in the {\it environment}, in which it finds itself". The dependence of the
variation rate on surrounding is common for biological as well as social
populations, for which the environment is given by the overall system, where
the population exists \cite{Hannan,Weidlich}.

The rate can become purely intrinsic only for an imaginary species called
Darwinian Demon, for which the rate is infinite \cite{Law}. But for any real
population, the rate always essentially depends on the environment, where the
population exists.

Moreover, the adjustment of each species to their surrounding by varying their
fitness, i.e. the rate, is a basic pillar of the Darwin theory of natural
selection, since all {\it organisms tend to exhibit considerable adaptation to
their natural environments} \cite{18,Kimbrough}.

Thus, it is the generally accepted and the sole correct understanding that
the population rate of any real species necessarily depends on environment,
that is, on the system, where the population exists, including all properties
of this environment. The population rate cannot be purely intrinsic, except the
fantastic case of Darwinian Demon.

In our case, the role of the rates is played by the responses $A_i$ showing
whether the given fraction increases or decreases the system utility, that is,
the sign and value of each $A_i$ are unambiguously defined by the relative
influence of the fraction on the utility of the system as a whole. This
definition immediately follows form the system-theory holistic approach, as
explained above.

\subsection*{Specification of trait groups and utility relations}

The above evolution equations can be applied to an arbitrary number of
groups of different nature. In order to concretely demonstrate the validity
of the approach, we consider a society formed by four groups: cooperators,
defectors, regulators, and outsiders. These are defined as follows.

\begin{itemize}
\item
    $x_1$: {\it Cooperators}, who contribute to the society, working for
the society benefit. In a human society, the cooperators form the actively
working force producing the gross domestic product, because of which they can
also be called producers. In a biological system, cooperators can be associated
with healthy cells.

\item
    $x_2$: {\it Defectors}, who, voluntarily or involuntarily, do not
contribute to the society, and can exist only owing to the work of cooperators.
In a social system, the groups that benefit from the society support without
contributing are prisoners, pensioners, and unemployed people. In a biological
organism, defectors can be represented by ill cells. Of course, there are many
degrees interpolating between fully contributing  and non-contributing behaviors
and this classification is obviously made for the sake of simplification.

\item
    $x_3$: {\it Regulators}, who maintain order in the society and punish
defectors and harmful outsiders. In a human society, this role is played
by the police, the army, and the order enforcing bureaucracy. To support the
existence of regulators, the society and cooperators have to pay the necessary
costs. In a biological organism, regulators can correspond to the cells of the
immune system.

\item
    $x_4$: {\it Outsiders}, who also exploit the society, as defectors, but
with the difference that they enter the society from outside. These exploiting
harmful outsiders should not be confused with immigrants. The latter, as soon
as they become members of the society, separate into cooperators, defectors,
and regulators. The harmful outsiders could be compared with those terrorists
that are supported by external sources. Outsiders could also be interpreted as
foreign raiding groups or even invading armies. For biological organisms,
outsiders could be pathogens or viruses infecting the organism.
\end{itemize}

This classification uniquely determines the signs of the coefficients entering
the evolution equations (\ref{5}) as follows. We assume that the external influx
can exist only for the outsiders and is positive (net in-flux). The diagonal
terms $A_{ii}$ can describe mutual competition or cooperation. It is negative
in the former case and positive in the latter case. In the presence of a dominant
mutual cooperation, the positivity of the diagonal terms $A_{ii}$ entails the
unbounded increase of the population of the $i$-group \cite{60,61Y}.
But, any real system has only finite resources and it is necessary to take into
account competition mechanisms that will eventually dominate to limit population
growth. This means that the diagonal coefficient $A_{ii}$, in general, could be
described as a function of the population fraction $x_i$, such that it would be
positive for small $x_i \ll 1$ and changing its sign when $x_i$ becomes larger.
However, there is the simpler way to capture this phenomenon without introducing
the sign change or higher-order terms with negative coefficients. This can be
done by always considering the diagonal terms $A_{ii}$ as being negative. Since
these terms enter the equations being factorized with $x_i^2$, at small
$x_i \ll 1$, such terms $A_{ii} x_i^2$ do not play an essential role, as compared
to the linear terms $A_i x_i$. And at larger $x_i$, the former terms makes it
impossible an infinite growth of populations.

In this way, keeping the diagonal terms negative provides mathematically the
same effect as if they would change the sign. Moreover, the existence of
competition between members of a group is a general phenomenon. This is natural even
for cooperators, not talking about other groups. Really, when the number of
cooperators would grow without limits, overpassing the available resources, they
would certainly start competing with each other for subsistence. This does not contradict the fact
that they do produce resources for the society. Real cooperators, or producers, are
always of dual nature, from one side, cooperating in the process of production, but,
from the other side, competing for the products.

The signs of all other coefficients are prescribed by the related utilities.
Summarizing, we have the following signs.

   {\it Cooperators}:
$$
A_1 > 0 \; , \qquad A_{11} < 0 \; , \qquad A_{12} < 0 \; ,
$$
\be
\label{6}
A_{13} < 0 \; , \qquad A_{14} < 0 \; , \qquad \Phi_1 = 0 \;   .
\ee
Cooperators are useful for the society, hence $A_1 > 0$. Defectors are not
useful for cooperators, hence $A_{12} < 0$. Regulators require the support of
cooperators which is a cost to cooperators, which defines $A_{13} < 0$.
And harmful insiders also are not useful for cooperators, that is, $A_{14} < 0$.

   {\it Defectors}:
$$
A_2 < 0 \; , \qquad A_{21} > 0 \; , \qquad A_{22} < 0 \; ,
$$
\be
\label{7}
 A_{23} < 0 \; , \qquad A_{24} < 0 \; , \qquad \Phi_2 = 0 \; .
\ee
The meaning of the chosen signs is again straightforward. Defectors are not
useful for the society, therefore $A_2 < 0$. Cooperators are necessary for
defectors who live at the expense of the former, hence $A_{21} > 0$. Regulators
suppress and punish defectors, that is, $A_{23} < 0$. Invaders are not useful
for defectors, so that $A_{24} < 0$.

   {\it Regulators}:
$$
A_3 < 0 \; , \qquad A_{31} > 0 \; , \qquad A_{32} > 0 \; ,
$$
\be
\label{8}
 A_{33} < 0 \; , \qquad A_{34} > 0 \; , \qquad \Phi_3 = 0 \;  .
\ee
Society needs to support regulators at a cost, thus $A_3 < 0$.
Cooperators, contributing to the society, are necessary for regulators,
hence $A_{31} > 0$. The role of regulators is to maintain order and to punish
defectors, whose presence justifies the existence of regulators, resulting
in $A_{32} > 0$. Similarly, regulators suppress harmful invaders, which
justifies the existence of regulators, giving $A_{34} > 0$.

   {\it Outsiders}:
$$
A_4 < 0 \; , \qquad A_{41} > 0 \; , \qquad A_{42} > 0 \; ,
$$
\be
\label{9}
A_{43} < 0 \; , \qquad A_{44} < 0 \; , \qquad \Phi_4 \geq 0 \;  .
\ee
Harmful outsiders are not useful for the society, which means $A_4 < 0$.
But cooperators are necessary for outsiders, hence $A_{41} > 0$. To some
extent, outsiders exploit defectors by taking a part of their share,
and benefit from their presence, hence $A_{42} > 0$. And, of course,
regulators, suppressing outsiders, are not useful to them, which implies
$A_{43} < 0$.   Only outsiders are here considered to contribute an influx
$\Phi_4$ to the system.

Let us emphasize that all coefficients above are always treated as the
corresponding utilities, or more precisely, as utility rates. And the
variables $x_i$ everywhere are the group population fractions defined in Eq. (2).

In specifying the relations between different groups, we should emphasize the
important symmetry principle \cite{14,15},
as a mechanism to ensure the structural stability of the dynamical system.
According to this principle, each term containing $A_{ij}$, entering the
system of equations, must have its counterpart containing $A_{ji}$. This principle
preserves the action-counteraction symmetry that is responsible for the system
structural stability. Note that the symmetry principle does not impose that
the matrix  $\{A_{ij}\}$ is symmetric, but just that if $A_{ij} \neq 0$, then
so is $A_{ji} \neq 0$. We are not aware of rigorous theorems proving the above
claim, but we have observed its relevance by developing a large number of
numerical simulations of multi-dimensional dynamical systems \cite{14,15}. In
the construction of the dynamical systems presented here, we use this structural
symmetry principle.

It is worth stressing that the signs of the coefficients are defined by
the utility relations of groups with respect to each other and with respect
to the whole society, but not by the interactions between the members of
different groups. Though, in many cases, the signs of the interactions
may be the same as those defined by utilities, but this is not always so. For
instance, the interaction of cooperators with defectors could produce
positive terms proportional to $N_1N_2$ in the equation for regulators, implying
that this interaction should lead to the increase of the numbers of regulators.
Or, some defectors, and even outsiders, after dealing with regulators, could
become cooperators, which would require to consider positive terms proportional
to $N_2N_3$ and $N_4N_3$ in the cooperator dynamics. It is possible to enumerate
a variety of such admissible interactions. If we would treat the evolution
equations as standard rate equations, whose coefficients are prescribed by
interactions, this would make the description not merely overcomplicated, but
also not well defined, because of the great number of all possible interactions
that could be taken into account. In addition, such terms, describing various
possible interactions, could break the action-counteraction symmetry of the system,
making it structurally unstable. Similarly, the coefficient $A_i$ characterizes
not the birth-death rate of the related population but the utility of the
corresponding $i$-th group for the society. This is why $A_1$ is positive because
of the usefulness of cooperators for the society, while $A_2$ and $A_4$ are
negative, since defectors and harmful outsiders exploit the society. Regulators
do not produce goods that would contribute to the society, but require costly
support from it, hence $A_3$ is negative.

The above classification is summarized by the following utility rate evolution
equations, in which the signs of the coefficients are explicitly taken into account.
As a result, we come to the following equations for the cooperators,
\be
\label{10}
\frac{dx_1}{dt} = \left ( A_1 - |A_{11}| x_1 -
|A_{12}| x_2 - |A_{13}| x_3 - |A_{14}| x_4 \right ) x_1 \; ,
\ee
for defectors,
\be
\label{11}
\frac{dx_2}{dt} = \left ( - |A_2| +  A_{21} x_1 -
|A_{22}| x_2 - |A_{23}| x_3 - |A_{24}| x_4  \right ) x_2 \; ,
\ee
for regulators
\be
\label{12}
\frac{dx_3}{dt} = \left ( - |A_3| +  A_{31} x_1 +
A_{32} x_2 - |A_{33}| x_3 + A_{34} x_4  \right ) x_3 \; ,
\ee
and for the outsiders,
\be
\label{13}
 \frac{dx_4}{dt} = \left ( - |A_4| +  A_{41} x_1 + A_{42} x_2 -
|A_{43}| x_3 - |A_{44}| x_4 \right ) x_4 + \Phi_4 \;  .
\ee
All coefficients not bracketed by an absolute value are positive.

\subsection*{Definition of relative utilities}

The dimensionless fractions, corresponding to trait groups, are defined
as normalized with respect to the total number of all group agents
$N_0 = N_1(0) + N_2(0) + N_3(0) + N_4(0)$ at the initial time $t = 0$.
Therefore, the initial conditions for the group population fractions $x_i(0)$
satisfy the equality
\be
\label{14}
 \sum_i x_i(0) = 1 \;  .
\ee
But the values $x_i(t)$, as functions of time $t$, are not generally
restricted to a simplex, i.e., their sum does not necessarily remain exactly
equal to $1$.

A priori, Eqs. (10) to (13) look rather complicated, containing 20
unknown parameters representing relative utilities that need to be defined.
Strictly speaking, it is possible to consider different situations, defining
the parameters according to particular situations. Below, we suggest a
general method leading to explicit relations between the parameters, thus
essentially decreasing their number.

First, let us recall that the diagonal quantity $A_{ii}$ has been
introduced in order to describe the competition between members of the same
$i$-group. This is the competition for the available resources allocated
for the particular group. On the other hand, the cost/benefit of an $i$-group
to the society is given by $A_i$. Hence, this group competes for the resources
$A_i$ allocated to it. In other words, the amount of resources $A_i$ is
proportional to the group carrying capacity. This implies that the diagonal
term $A_{ii}$, describing the group internal competition, should be taken as
anti-correlated with, or more generally varying inversely to, the carrying
capacity $|A_i|$. A convenient form is to take $A_{ii}$ as being inversely
proportional to $|A_i|$, so that
$$
 |A_{ii}| = \frac{C}{ |A_i| } \;  ,
$$
where $C$ is a positive constant.

In a self-organized society, the available resources are those that are
produced by cooperators, who are the sole net contributors to the society
in term of resources and who compete for these resources. This is equivalent
to saying that
\be
\label{15}
| A_{11} | = A_1 \;   .
\ee
Then, comparing the above two equalities, we get the constant $C = A_1^2$.
From here, we obtain the general relation for the diagonal terms
\be
\label{16}
 | A_{ii} | = \frac{A_1^2}{ | A_i| } \;  .
\ee

Nondiagonal terms $A_{ij}$ characterize mutual utilities of groups with
respect to each other. There exist several expressions for mutual utilities
\cite{Aco,Mou}, among which the most symmetric one is the
Bernoulli-Nash mutual utility
\be
\label{17}
  | A_{ij} | = \sqrt{ | A_i A_j | } \; ,
\ee
which we adopt in what follows.

The Bernoulli-Nash mutual utility relates the mutual
utility of different groups through their utility to the society to which
they all belong. Different groups influence each other by influencing the
society. For instance, if the $i$-group is not useful to the society, so that
$A_i$ is zero, then this group is certainly not useful for the $j$-group, since
the latter is part of the same society.
The geometric mean for characterizing
the mutual utility automatically ensures these properties and
additionally conserves the dimensionality of all terms
involved. Note that this form of the mutual utility, called the
Bernoulli-Nash utility, is often used in various applications \cite{Aco,Mou}.

Let us introduce the dimensionless parameters
\be
\label{18}
  a \equiv \frac{|A_2|}{A_1} \; , \qquad  b \equiv \frac{|A_3|}{A_1} \; ,
\qquad c \equiv \frac{|A_4|}{A_1} \;  ,
\ee
quantifying the resources consumed by the corresponding groups, and the
dimensionless outsider influx
\be
\label{19}
\vp \equiv \frac{\Phi_4}{A_1N_0} \; ,
\ee
all of which are non-negative quantities. And let us measure time in units
of $1/A_1$. Then Eqs. (10) to (13) reduce to the system of equations
\be
\label{20}
\frac{dx_i}{dt} = f_i(x_1,x_2,x_3,x_4 ) \qquad ( i = 1,2,3,4)  \; ,
\ee
with the right-hand sides
\be
f_1 = \left ( 1 - x_1 - \sqrt{a} \; x_2 - \sqrt{b} \; x_3 - \sqrt{c} \; x_4
\right ) x_1 \; ,
\label{f1}
\ee
\be
f_2 = \left ( -a + \sqrt{a} \; x_1 - \frac{1}{a}\; x_2 - \sqrt{ab} \; x_3 -
\sqrt{ac} \; x_4 \right ) x_2 \; ,
\label{f2}
\ee
\be
f_3 = \left ( -b + \sqrt{b} \; x_1  + \sqrt{ab} \; x_2  - \frac{1}{b}\; x_3 +
\sqrt{bc} \; x_4 \right ) x_3 \; ,
\label{f3}
\ee
\be
f_4 = \left ( -c + \sqrt{c} \; x_1  + \sqrt{ac} \; x_2  - \sqrt{bc} \; x_3 -
\frac{1}{c}\; x_4  \right ) x_4 + \vp \;  .
\label{f4}
\ee

These equations are assumed to be complemented by the corresponding initial
conditions $x_i(0)$. By definition, the fractions $x_i$ are non-negative.
Therefore, we shall be looking only for non-negative solutions of the
evolution equations (\ref{20}) with (\ref{f1}-\ref{f4}).

\section*{Evolutionally stable strategies}

\subsection*{Single-group evolution}

Let us start the analysis with the case where, at the initial time, there is
just one group and other group fractions are set to zero. Suppose,
first, that there exists just the group of cooperators. It follows from
Eq. (\ref{20}) with (\ref{f1}) that the cooperator population is
described by the logistic equation
$$
\frac{dx_1}{dt} = (1 - x_1) x_1 \; ,
$$
whose solution is
$$
x_1(t) = \frac{x_1(0) e^t}{1 + x_1(0) (e^t - 1)} \; .
$$
Substituting here the initial condition $x_1(0) = 1$ yields $x_1(t) = 1$
for all times $t$. This is clear since the initial condition coincides with the stable
fixed point $x_1^* = 1$. The meaning of this solution is
that cooperators can perfectly exist without other groups.

Let us now consider the population of any other group, except cooperators,
as existing alone at the initial time. For instance, let us take the equation
for defectors,
$$
\frac{dx_2}{dt} = - (a + \frac{1}{a} x_2) x_2 \; ,
$$
with the initial condition $x_2(0) = 1$. The solution is
$$
x_2 = \frac{a^2}{1 + (a^2 - 1) e^{at}}.
$$
This population decays to zero with time. This is easy to understand as far
as defectors do not produce but merely consume and cannot survive being left
alone. The same result holds for other groups, i.e., for $x_3$ as well as for $x_4$,
which cannot survive without cooperators.

\subsection*{Coexistence of cooperators and defectors}

More interesting is the case of two coexisting groups, cooperators and
defectors. Actually, this is the classical situation from which one
usually starts analyzing group interactions. In such a case, Eqs. (\ref{20})
reduce to two equations
\be
\label{21}
 \frac{dx_1}{dt} = \left ( 1 - x_1 - \sqrt{a} \; x_2 \right ) x_1 \; ,
\qquad
\frac{dx_2}{dt} = \left ( -a + \sqrt{a} \; x_1 - \frac{1}{a}\; x_2
\right ) x_2 \;  .
\ee

This system possesses two evolutionary stable states. One is given by
the stationary solution
\be
\label{22}
  x_1^* = 1 \; , \qquad x_2^* = 0
\ee
that is stable when
\be
\label{23}
 a > 1 \;  .
\ee
The meaning of this state is evident: if defectors attempt to consume more
than cooperators can produce, the stable state can exist only when the
very greedy defectors are absent, and only cooperators live.

Another evolutionary stable state is
\be
\label{24}
 x_1^* = \frac{1+a^{5/2}}{1+a^2} \; , \qquad
x_2^* = \frac{1-\sqrt{a}}{1+a^2}\; a^{3/2} \;  ,
\ee
being stable under the condition
\be
\label{25}
0 \leq a \leq 1 \;  .
\ee
The meaning of this state is again rather clear. When defectors consume
only a part of what is produced by cooperators, they can coexist with them.
The dependence of the evolutionary stable population fractions
$x_1^*$ and $x_2^*$, on the amount of the resources $a$ consumed by defectors,
is shown in Fig. 1. The minimal fraction of cooperators happens when
\be
\label{26}
 \min_a x_1^*(a) = 0.940 \qquad ( a = 0.565) \;  ,
\ee
while defectors have a maximum
\be
\label{27}
 \max_a x_2^*(a) = 0.083 \qquad ( a = 0.480 ) \;  .
\ee
These minimum and maximum do not coincide. In a stable society consisting of
cooperators and defectors, the fraction of defectors cannot overpass the limit
of order $10\%$.

These logical conclusions are drastically different from the result of
the replicator equation for the case of two coexisting groups
(cooperators and defectors), which predicts that the sole stable state consists
solely  of defectors, without any cooperators. Such a state is obviously
impossible in a self-organized society since, without cooperators, nothing is
produced and defectors would have no means for survival.

\subsection*{Coexistence of three groups}

Let us now consider a society formed by three trait groups, cooperators,
defectors, and regulators. The evolution equations (\ref{20}) form a
three-dimensional dynamical system
$$
\frac{dx_1}{dt} = \left ( 1 - x_1 - \sqrt{a} \; x_2 - \sqrt{b} \; x_3
\right ) x_1 \; ,
$$
$$
\frac{dx_2}{dt} = \left ( -a + \sqrt{a} \; x_1 - \frac{1}{a}\; x_2 -
\sqrt{ab} \; x_3 \right ) x_2 \; ,
$$
\be
\label{28}
 \frac{dx_3}{dt} = \left ( -b + \sqrt{b} \; x_1  + \sqrt{ab} \; x_2  -
\frac{1}{b}\; x_3 \right ) x_3 \; .
\ee
Looking for stationary solutions, we select positive and stable fixed
points, employing the standard Lyapunov analysis. The system possesses
four types of evolutionary stable solutions.

The first stable state is given by the fixed point
$$
x_1^* = \frac{1+a^2b^2+a^{5/2}(1+b^2)+b^{5/2}(1-a^2)}{(1+a^2)(1+b^2)} \; ,
$$
\be
\label{29}
x_2^* = \frac{1-\sqrt{a}(1+b^2)-b^2(1-2\sqrt{b})}{(1+a^2)(1+b^2)}\; a^{3/2} \; ,
\qquad
x_3^* = \frac{1-\sqrt{b}}{1+b^2}\; b^{3/2} \; ,
\ee
which is stable under the conditions
\be
\label{30}
0 \leq a < \left (\frac{2b^2\sqrt{b}-b^2+1}{1+b^2} \right )^2 \; , \qquad
0 \leq b < 1 \;   .
\ee
This stable state occurs when defectors and regulators consume reasonable amount
of the resources of the society, so that all three groups can coexist.

The second stable state corresponds to the solution
\be
\label{31}
x_1^* = \frac{1+b^{5/2}}{1+b^2} \; , \qquad  x_2^* =  0\; , \qquad
x_3^* = \frac{1-\sqrt{b}}{1+b^2}\; b^{3/2} \;   ,
\ee
being stable when
\be
\label{32}
a \geq \left (\frac{2b^2\sqrt{b}-b^2+1}{1+b^2} \right )^2 \; ,
\qquad 0 \leq b < 1 \;   .
\ee
Here, the defectors, who consume too much die out, while cooperators
and regulators coexist with each other.

The third stationary solution is
\be
\label{33}
x_1^* = \frac{1+a^{5/2}}{1+a^2} \; , \qquad
x_2^* =  \frac{1-\sqrt{a}}{1+a^2}\; a^{3/2} \; , \qquad x_3^* = 0    ,
\ee
which is stable if
\be
\label{34}
0 < a < 1 \; , \qquad b > 1 \;  .
\ee
In this case, the excessive consumption (or cost) of regulators makes them
unprofitable for the society, so that they are suppressed to zero, while
defectors who are not too greedy remain viable at an optimal finite
fraction $x_2^*$.

Finally, the fourth state is described by the set
\be
\label{35}
x_1^* = 1 \; , \qquad x_2^* = 0 \; , \qquad x_3^* = 0    ~,
\ee
which is stable provided that
\be
\label{36}
a \geq 1 \; , \qquad b \geq 1 \;    .
\ee
The meaning of this solution is again clear. Defectors and regulators,
trying to consume more than what the cooperators produce constitute
an unsustainable burden for the society, which prefers to eliminate them.

The phase portrait for the case of three coexisting groups is shown in Fig. 2.
Figure 3 shows the dependence of the evolutionary stable cooperator
fraction $x_1^*$ as a function of the amount of the resources consumed by
defectors $(a)$ and regulators $(b)$. The function $x_1^*$ has a global minimum
\be
\label{37}
 \min_{a,b} x_1^* = 0.905 \qquad
( a = 0. 456 \; , \;\; b= 0.565 ) \; .
\ee
The evolutionary stable fraction of defectors $x_2^*$ is shown in Fig. 4
as a function of $a$ and $b$. This function enjoys a global maximum
\be
\label{38}
\max_{a,b} x_2^* = 0.083 \qquad
( a = 0. 480 ) \;   ,
\ee
when either $b = 0$ or $b \geq 1$. The stable fraction of regulators
$x_3^*$ is presented in Fig. 5. This function has a global maximum
\be
\label{39}
 \max_{a,b} x_3^* = 0.083 \qquad
( a \geq 0 \; , \; \; b = 0.480 ) \;   .
\ee
These results show in particular that, in order to achieve an evolutionary
stable state in a self-organized society, the fractions of defectors and
regulators should not exceed about $10\%$ each.

\section*{Applications to biological and human societies}

\subsection*{Biological societies without defectors}

As follows from the above analysis, there can exist societies with no defectors.
A good example of this is a bee colony. A honey bee colony typically consists
of three kinds of adult bees: workers, drones, and a queen \cite{61}. Several
thousand worker bees cooperate in nest building,
food collection, and brood rearing. Each member has a definite task to perform,
related to its adult age. But surviving and reproducing take the combined efforts
of the entire colony. Individual bees (workers, drones, and queens) cannot survive
without the support of the colony. Even drones (males honey bees) are necessary
members of the colony. In that sense, all bees are cooperators, having no defectors.
They also do not have separate groups of regulators or defenders. Each bee perfectly
knows its task. And in the case of danger, all bees defend their hive. But, excluding
the cases of external aggression, in the normal situation, the bee colony is an
example of a society where practically all members are cooperators. As an extreme,
drones which are useful mainly in mating are expendable members of the colony,
driven from the colony as winter approaches where they perish from cold and
starvation. This mechanism removes defectors, in the sense defined in our model.

Another example of a biological society without defectors is a colony of ants.
The colony is typically divided into the following castes, or classes: queens
(reproductive females), males, and workers (nonreproductive females). Although
there are great variations in social structure among ant colonies, the basic
features are common to most species \cite{62}. There are no defectors in an
ant colony, all ants accomplish a job contributed to the whole group. Often,
one classifies the workers into two castes, minor workers and major workers,
or soldiers. However, this is done on the basis of the difference in their
sizes, but does not mean that the sole job of soldiers is defending the nest,
as they do accomplish other jobs \cite{63,64,65}. All workers participate
in the defense when necessary, though soldiers may do this more often. But
soldiers also take part in other numerous acts, such as foraging food, taking
care of larva, carrying dead nestmates, and so on \cite{66,67,62,Mir}.
So, strictly speaking, soldiers are just bigger working ants. For this reason,
they also are called foragers since, being bigger, they can carry heavier
pieces of food.

In order to perform a quantitative comparison with our theory, we consider
the ratio of the number of foragers ($N_F$) to the number of other working ants
($N_W$), which is termed the cast ratio, $r \equiv N_F / N_W$. The fraction of
foragers with respect to the total population is then
\be
 \frac{N_F}{N_F + N_W} = \frac{r}{r+1} \;  .
\ee
Defense is a part of the labor repertory, which can be denoted as $\delta$. Then,
the fraction of the defense trait reads as
\be
\label{A1}
  x_3 = \frac{r}{1+r} \; \delta \; .
\ee
Recall that defenders are a part of the group of regulators.

The caste ratio of ants has been investigated in numerous articles (see the
literature cited above and references therein). This ratio can vary in a wide
range, depending on the colony age, time of the season, external danger, climate,
and so on. To get estimates, it is reasonable to consider average numbers. Thus,
the average caste ratio, measured for 503 colonies by Kaspari and Byrne \cite{64},
was found to be $r = 0.11$, which gives the fraction of foragers $0.099$. The
repertory of ant labor has been intensively studied by Wilson \cite{66,67} and
Oster and Wilson \cite{62}, who define the fraction of the defense behavior of
foragers, among the total number of their acts, as $\delta = 0.345$. These data
give the fraction of the defense trait (\ref{A1}) as $x_3 = 0.034$. This value
is nicely compatible with our estimate requiring that the maximal fraction of
regulators in stable equilibrium should not be larger than $x_3^* = 0.08$.

One can thus consider an ant colony as a society with two trait groups,
cooperators and regulators (soldiers). It seems that defectors rarely occur
among social insects, such as bees, wasps, ants and termites.
This is probably due to the strong selection pressure
acting on these specific biological species, so that costly defectors are selected out. Defense
is often necessary and one does find a non-zero defense trait group in an ant
colony. But defense is in this case principally directed to external aggressions,
in particular against other foreign ants for ant colonies, and occupies a rather small fraction
of the population, as predicted by our model.

\subsection*{Classification by distance from equilibrium}

In real life, no society can be considered to be in an absolute stable equilibrium.
Hence, we do not expect any society to lie exactly at or even close to any of the
equilibrium points described above, either because of the previous history or
because of external influences that may prevent the dynamics to have sufficient
time to converge to the corresponding equilibrium point. How can we then
characterize the relative stabilities of different societies and distinguish
between them?

To answer this question, we propose to introduce the {\it distance from equilibrium}
\be
\label{40}
 D(t) \equiv \sqrt{(x_1(t)-x_1^*)^2 + (x_2(t)-x_2^*)^2+(x_3(t)-x_3^*)^2 } \;   ,
\ee
and use this metric to classify societies in terms of their relative stability,
according to whether they are further or closer to an equilibrium state.

\subsection*{Application to human populations}

In order to show that the proposed approach yields not merely qualitatively
reasonable results, but also provides a realistic quantitative picture,
let us consider the social systems represented by populations of different
countries. Then, the fraction of cooperators $x_1$ describes the active working
part of the society, producing all resources. As defectors ($x_2$), we count the
fraction that does not take a direct part in the production of goods, but is
supported by the society; these are unemployed people, pensioners, and prisoners.
Regulators ($x_3$) are police, army, and law-enforcing bureaucrats.

To define the parameters $a$ and $b$, we resort to the analogy between human
and biological societies. These parameters are defined in Eq. (\ref{18}) through
the quantities $A_i$ that symbolize the utilities of the related groups.
Recall that utility for a human society is equivalent to fitness for a biological
society. In biology, fitness is defined as the product of viability and fecundity
rate \cite{52,53,54,55}. Viability is the capacity for survival of a group \cite{55},
while the fecundity rate or potential reproductive capacity rate \cite{52} is the
number of offsprings produced by an organism per unit time.

Similarly, we define the utility $A_i$ as the product $A_i = C_i G_i$ of the
capacity $C_i$ available to a group for its survival and of the growth rate $G_i$
characterizing the relative group growth or its potential growth due to the
increasing capacity. In human societies, the growth of a group is usually
proportional to the increase of the resources provided to this group
\cite{68,69,70,71}. Conversely, increasing the group resources provides means
for the potential group growth. In this way, the parameters $a$ and $b$ are defined
as the ratios
\be
\label{41}
a \equiv \frac{ |A_2|}{A_1} = \frac{ |C_2|G_2}{|C_1|G_1}  \; , \qquad
 b \equiv \frac{ |A_3|}{A_1} = \frac{ |C_3|G_3}{|C_1|G_1}  \;.
\ee

In more or less stable countries, the growth rate of the total resources
produced by cooperators is assumed to coincide with the growth rates of
the resources available for each group. Then, the system parameters become
$a = |C_2|/C_1$ and $b = |C_3|/C_1$. This is probably not always the case,
and some groups can grow faster than others. An example could be the
bureaucracy in Russia, which grows much faster than the country GDP. Another
example may be the accelerated growth of the weight in percentage of GDP of
government spending in western countries, including the US, as a reaction to the
great recession that started in 2008. In many countries, the growth of spending
programs, such as in Japan in the mid-1990s, or in health care, also grow faster
than GDP. However here, for simplicity, we assume that the growth rates of the
different groups are proportional to GDP, leaving the fine structure related
to the possible unbalance between the growth rates for future economical
investigations.

Within this assumption leading to $a = |C_2|/C_1$ and $b = |C_3|/C_1$,
the parameter $a$ characterizes the portion of the resources, that is
of the budget, the society pays for supporting the defectors, while $b$ is
the portion of the budget spent for the regulators. All these data can be
found  on the websites provided at the end of the list of References.

Note that different Internet sources, even the official ones, quite often give
different information on a country total population, labor force, employment,
number of military or police forces, and so on. We have used the data for 2008,
since only those data were presented in the most complete and uniform way for
the chosen countries. In those cases where we could not find the appropriate
data for 2008, we used the information for the nearest available year. It was
possible to do so because there were no sharp changes in numbers for adjacent
years. The differences in data should provide only small deviations in initial
conditions or parameters and, because of the structural stability of the
system (\ref{28}), it should not drastically change the results. For example,
to estimate the number of retired people in France, we used the data from the
population pyramid of France for 2010, instead for 2008, to be able to take
into account early retirements.

We do not expect any country to lie exactly at or even close to any of the
equilibrium points described above, because there always exist changes of
conditions at the geopolitical and economic levels that may prevent the
dynamics to have sufficient time to converge to the corresponding equilibrium
point. We use the distance from equilibrium (\ref{40}) as the metric to
classify the societies in terms of their relative stability, according to
whether they are further or closer to an equilibrium state.

The initial values and parameters entering the system of equations (\ref{28})
are taken from the data characterizing the corresponding countries. All these
data can be found on the websites \cite{ws1,ws2}. For example, we illustrate below
the case of Israel \cite{ws1}, explaining how the data, characterizing the considered
countries, have been estimated.

The total population of Israel in 2008 is estimated as $7.34$ M. According to
our definition, defectors are assumed to constitute the part of the population
that does not contribute to the society, but instead is supported by it. These
are pensioners, unemployed, and prisoners. Pensioners compose $9.8\%$ of
the total population.

The labor force of Israel in 2008 is estimated as 2.96 M people.
Unemployment rate is $6.1\%$ of its labor force, which equals $0.18$ M of the
total population of Israel. Hence, the percentage of unemployed with
respect to the total population is $2.46\%$.

The number of prisoners in Israel in 2008 is 0.013 M, which gives $0.177\%$
of the total population. Thus the percentage of defectors is evaluated, in total,
as $9.8\% + 2.46\% + 0.177\% = 12.4\%$, which gives $x_2 = 0.124$.

The group of regulators is formed by the army, police, and bureaucrats.
The army is strong of $0.185$ M persons and the police has a force of $0.035$ M
people. This gives the total number of the population in the army and police
as $0.22$ M. In other terms, the army and police compose $3\%$ of the total
population.

The number of bureaucrats makes $16.5\%$ of the labour force of $2.96$ M.
This gives the total number of bureaucrats in the country as $0.488$ M.
So, the percentage of bureaucrats with respect to the total population
is $6.65\%$ . Therefore the fraction of regulators $x_3$ is evaluated as
$6.65\% + 3\% = 9.65\%$, which gives $x_3=0.097$.

The fraction of {\it cooperators}, $x_1$, at the given initial time, is
calculated as $x_1 = 1 - (x_2 + x_3) = 1 - (0.124 + 0.097) = 0.779$.
Note that this is different from the employed productive fraction, as we
include all the children, who arguably grow human capital, and people working
at home (housewives for instance) who do not appear in the unemployed statistics
but nevertheless contribute to the production of the country.

The parameter $a$ is the cost of the group $x_2$ for the group $x_1$,
and the parameter $b$ is the cost of the group $x_3$ for the group $x_1$.
We consider $a$ as the part of the budget spent on {\it social protection}.
In 2008, Israel spent on social protection $25.5\%$ of the budget,
which means that $a=0.255$.

On {\it national defense, public order and safety}, and {\it state governing},
Israel spent respectively $16.4\%$, $3.8\%$, and $12.7\%$ of its total budget.
In total, it thus gives $32.9\%$, hence $b=0.329$.

In this way, analyzing the system of equation (\ref{28}) for Israel, we employ
the found parameter values $a = 0.255$ and $b = 0.329$, and take as initial
conditions $x_1(0) = 0.779, x_2(0) = 0.124, x_3(0) = 0.097$.

We have followed the same procedure for the other nine countries. All data
are taken from the websites \cite{ws2}.

As shown in the Table 1, we find for the ten analyzed countries that they all lie
in the domain I of the phase portrait depicted in Fig. 2, for which there is a
stable fixed point with a finite fraction of all three groups of cooperators,
defectors and regulators. This is not surprising, since the common structure of
all countries always includes these three groups. We are not aware of countries
that would not contain some of these groups. This is in contrast to social insect
colonies, for which defectors are too costly and rarely exist, as discussed for the
cases of bee and ant colonies.

The real value of the metric (\ref{40}) rests in the possibility to quantify how
far from their corresponding equilibrium point is each country, given its
corresponding loss $a$ due to defectors and cost $b$ of regulators.
In order to understand the meaning of the Table 1, one could interpret
the coefficients $a$ and $b$ for each country as reflecting the choice
of the society, for instance through evolution and/or a political process.
In other words, we interpret the parameters $a$ and $b$ measured for
each country as the independent variables determined by the society.
Then, given the set $a$ and $b$, our theory predicts what should be the optimal
fraction of the three groups of cooperators, defectors and regulators, where
optimality is referring to maximum stability. Another possible interpretation
is that the distance $D$ defined by (\ref{40}) for each country quantifies a
prediction of a future probable evolution of the three groups, if the relative
losses $a$ associated with defectors and costs $b$ of regulators remain unchanged.
Indeed, being at a distance to the stable fixed point, a given country is expected
to see its population partition in the three groups to evolve towards the fractions
given by the values of the stable fixed points for the measured $a$ and $b$. We are,
of course aware that this may not happen, since the parameters $a$ and $b$ can be
varied because of political reasons and technology changes.

Let us mention that it could be possible to consider the deviations from the
stationary states for each of the variables $x_1, x_2, x_3$ separately. This
would show the relative instability of the country with respect to the particular
characteristics. However, the overall stability is connected with all three
deviations, since it is possible to compensate one large deviation by other
deviations. Although the consideration of partial separate deviations is admissible,
here we limit ourselves to the analysis of the global deviation $D(t)$.

Some interesting features are worth noting. For all countries, there are not
sufficiently many cooperators, too many defectors and too many regulators,
compared to the normative values associated with the stable equilibrium points.
This suggests that societies choose to allocate their resources in a way that is
suboptimal with respect to stability, considering other factors such as social
preferences as well as historical culture, a hardly surprising observation. As
a consequence, the possibility for changes and instabilities in the presence of
external perturbations should not be underestimated, since it results from the
fundamental choice taken consciously or through the force of history to function
far from the stability point.

It may be surprising to find Israel at the top of the ranking. One could advance
the interpretation that its special geopolitical situation has forced it to evolve
closer to a regime of structural stability. Other countries in the West may have,
at least for a while, the luxury to choose less stable policies.

Another interesting point is made by comparing Switzerland with USA and Spain.
While the cost of defectors (main social and retirement costs) is much
higher in Switzerland than in the USA ($a=0.407$ compared with $0.19$),
Switzerland is ranked higher (more stable) than the USA because
it has a much lower cost for regulators ($b=0.197$ compared with $0.304$
reflecting its much smaller relative military budget), confirmed by its smaller
fraction of regulators. In contrast, Spain has lower costs from defectors and
for regulators than Switzerland, but more defectors and almost twice as many
regulators, hence its relatively much lower rank.

The laggard of the list is Russia. Examining its parameters $(a, b)$ and its
three group fractions $(x_1, x_2, x_3)$, it is clear that its woes are rooted
in the combination of having the smallest number of cooperators and the largest
number of defectors, as well as the largest number of regulators. This points
clearly to the general directions of reforms that could improve its state.

Finally, it is quite remarkable that the classification of countries presented
in the Table, in terms of their distance from equilibrium, is in excellent
agreement with the general understanding about their respective economic stability.
It is particularly interesting that, for the European countries, the ranking
shown in the Table coincides with the ranking in terms of GDP per capita given
by the International Monetary Fund, which can be found on the site \cite{ws3}.

\subsection*{Dynamics in presence of noise}

Real societies can never reach their absolute equilibrium states because of the
everlasting existence of external perturbations, which can be modeled as a kind
of noise superimposed on the deterministic dynamics that we have considered until
now. To investigate the consequence of such a noise, we have studied the behavior
of the population fractions as functions of time in the presence of random
perturbations that are modeled by adding to the evolution equations (\ref{20})
Gaussian white noise characterized by the standard deviation $\sigma$. Then,
each population fraction obeys the equation
\be
x_i(t_n + \Dlt t) - x_i (t_n) = f_i(t_n) \Dlt t +
\sgm \sqrt{\Dlt t}\; R(t_n) \;  ,
\ee
where $R(t_n)$ are random numbers, $\Delta t$, a small time interval,
$t_n = n \Delta t$, and $n = 1,2,\ldots$.

For illustration, we take the data corresponding to Israel. The resulting
temporal behavior of group fractions is shown in Figs. 6 to 8. From these
figures, it is seen that, in the presence of external perturbations, even
a rather stable country, such as Israel, from time to time may experience
quite strong deviations far from equilibrium.

\section*{Conclusion}

A novel approach has been suggested for describing the evolution of
{\it self-organized} structured societies composed of several trait groups.
The main idea is to consider the utility rate equations, whose parameters are
characterized by the utility of each group with respect to the society as
a whole and by the mutual utilities of groups with respect to each other.

The principal point is that we consider {\it self-organized} societies whose
resources are those produced by the societies themselves.

We have analyzed in detail the cases of two coexisting groups (cooperators
and defectors) and of three groups (cooperators, defectors, and regulators).
In a self-organized society, neither defectors nor regulators can overpass
the maximal fractions of order $10\%$. This is in contrast with the standard
replicator equation, where defectors can overpass the amount of cooperators,
if no regulators or punishers are present.

According to the studied equations, there can exist societies without defectors.
Examples of such societies are rather common for biological species. For
instance, there are no defectors in bee or ant colonies.

It is numerically demonstrated that the suggested approach, though being
relatively simple, gives reasonable results when applied to ten countries,
from Israel to Russia. We have shown in particular how the ranking of countries
in terms of the distance to their corresponding stable equilibrium point is
in remarkable concordance with more standard economic ranking based, for
instance, on GDP per capita.

A possible generalization of the evolution equations can be done by considering
the society behavior during long time periods, when the system parameters $a$
and $b$ become functions of time. For example, the amount of the resources
produced by cooperators in the long run can be taken as a function of time,
similarly to the time dependence of the carrying capacity studied in other
models \cite{60,72,73}. In contrast, as has been explained in the Introduction,
we have considered here the intermediate time scales lying between the short
times of the interactions among the group members and the large scale of the
system lifetime.

In the present paper, we have limited our considerations to the case of closed
societies, when there are no invaders from outside. For social systems,
such invaders could be associated with terrorists or foreign armies, invading
the assembly of cooperators, defectors, and regulators. For biological systems,
these invaders could correspond to viruses and pathogens penetrating a
biological organism formed of healthy cells (cooperators), ill cells (defectors),
and the cells of immune system (regulators). The role of such invaders will be
studied in a separate publication.

\section*{Acknowledgments}

We acknowledge financial support from the ETH Competence Center "Cooping with
Crises in Complex Socio-Economic Systems".

\newpage

\section*{Figure Legends}

{\parindent=0pt
\vskip 1cm
{\bf Fig. 1}. Evolutionary stable fractions of cooperators $x_1^*$ (solid line) and
defectors $x_2^*$ (dashed line) as functions of the relative amount of resources
$a$ consumed by defectors, in the case of the coexistence of two trait groups,
cooperators and defectors.

\vskip 1cm
{\bf Fig. 2}. Phase portrait for the evolutionary stable states in the case of three
coexisting trait groups, cooperators, defectors, and regulators on the plane
of the relative amount of resources consumed by defectors $(a)$ and regulators
$(b)$. In region I, all three groups coexist, having nonzero fractions. In
region II, there exist only cooperators and regulators, while there are no
defectors $(x_2^* = 0)$. In region III, cooperators and defectors coexist,
but there are no regulators $(x_3^* = 0)$. In region IV, only cooperators are
present, without either defectors or regulators $(x_2^* = 0, x_3^* = 0)$.

\vskip 1cm
{\bf Fig. 3}. Evolutionary stable fraction of cooperators $x_1^*$ as a function
of the parameters $a$ and $b$.

\vskip 1cm
{\bf Fig. 4}. Evolutionary stable fraction of defectors $x_2^*$ as a function of
the parameters $a$ and $b$.

\vskip 1cm
{\bf Fig. 5}. Evolutionary stable fraction of regulators $x_3^*$ as a function of
the parameters $a$ and $b$.

\vskip 1cm
{\bf Fig. 6}. Temporal behavior of the cooperator fraction $x_1$ in the presence
of noise of different intensity for the parameters $a = 0.26$, $b = 0.33$
corresponding to Israel: (a) $\sigma = 0$; (b) $\sigma = 0.03$; (c) $\sigma = 0.06$;
(d) $\sigma = 0.09$.

\vskip 1cm
{\bf Fig. 7}. Temporal behavior of the defector fraction $x_2$ in the presence
of noise of different intensity for the parameters $a = 0.26$, $b = 0.33$
corresponding to Israel: (a) $\sigma = 0$; (b) $\sigma = 0.03$; (c) $\sigma = 0.06$;
(d) $\sigma = 0.09$.

\vskip 1cm
{\bf Fig. 8}. Temporal behavior of the regulator fraction $x_3$ in the presence
of noise of different intensity for the parameters $a = 0.26$, $b = 0.33$
corresponding to Israel: (a) $\sigma = 0$; (b) $\sigma = 0.03$; (c) $\sigma = 0.06$;
(d) $\sigma = 0.09$.

}

\newpage

\section*{Tables}

\vskip 2cm

\begin{table}[!ht]
\caption{Data for different countries corresponding to the present fractions
of cooperators $x_1$, defectors $x_2$, and regulators $x_3$, compared to their
stable stationary solutions (given in brackets). The portion of the budget,
consumed by defectors is $a$ and the relative amount spent for regulators
is $b$. Countries are classified according to their distance $D$ from equilibrium
defined by expression (\ref{40}).
}
\vskip 3mm
\begin{tabular}{|l|c|c|c|c|c|c|} \hline
               & $D$    & $a$   & $b$   & $x_1\;\;(x_1^*)$ &  $x_2\;\;(x_2^*)$ & $x_3\;\;(x_3^*)$  \\ \hline
1. Israel      & 0.173  & 0.255 & 0.329 & 0.779\;\;(0.933) & 0.124\;\;(0.050)  & 0.097\;\;(0.073)  \\
2. Japan       & 0.203  & 0.350 & 0.192 & 0.761\;\;(0.940) & 0.157\;\;(0.068)  & 0.082\;\;(0.046)  \\
3. Switzerland & 0.217  & 0.407 & 0.197 & 0.750\;\;(0.934) & 0.186\;\;(0.071)  & 0.064\;\;(0.047)   \\
4. USA         & 0.242  & 0.190 & 0.304 & 0.740\;\;(0.945) & 0.167\;\;(0.039)  & 0.082\;\;(0.069) \\
5. Germany     & 0.282  & 0.451 & 0.196 & 0.704\;\;(0.919) & 0.240\;\;(0.072)  & 0.056\;\;(0.047)  \\
6. France      & 0.284  & 0.414 & 0.193 & 0.690\;\;(0.934) & 0.205\;\;(0.072)  & 0.105\;\;(0.046)  \\
7. Italy       & 0.286  & 0.385 & 0.250 & 0.697\;\;(0.929) & 0.233\;\;(0.067)  & 0.070\;\;(0.059)   \\
8. Spain       & 0.304  & 0.340 & 0.187 & 0.691\;\;(0.942) & 0.238\;\;(0.067)  & 0.071\;\;(0.044)  \\
9. Greece      & 0.314  & 0.365 & 0.294 & 0.665\;\;(0.926) & 0.234\;\;(0.063)  & 0.101\;\;(0.067)  \\
10. Russia     & 0.471  & 0.363 & 0.225 & 0.546\;\;(0.934) & 0.322\;\;(0.067)  & 0.132\;\;(0.053) \\ \hline
\end{tabular}

\label{tab:label}
\end{table}

\newpage

%Figura 1
\begin{figure}[ht]
\centerline{\includegraphics[width=10cm]{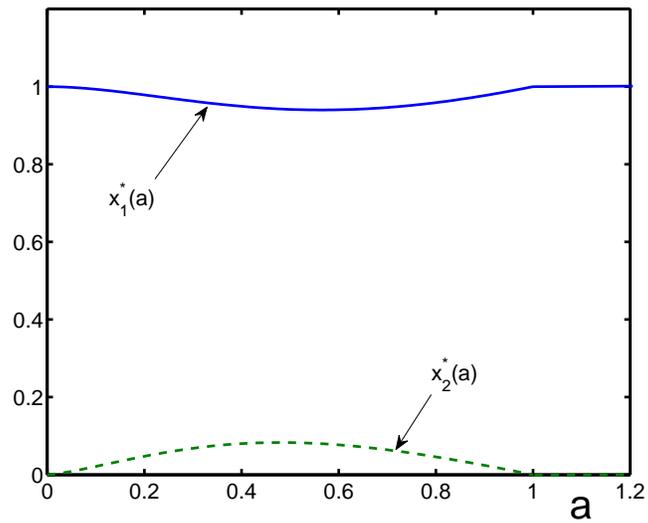} }
\caption{Evolutionary stable fractions of cooperators $x_1^*$ (solid line) and
defectors $x_2^*$ (dashed line) as functions of the relative amount of resources
$a$ consumed by defectors, in the case of the coexistence of two trait groups,
cooperators and defectors.}
\label{fig:Fig.1}
\end{figure}

\newpage

%Figura 2
\begin{figure}[ht]
\centerline{\includegraphics[width=12cm]{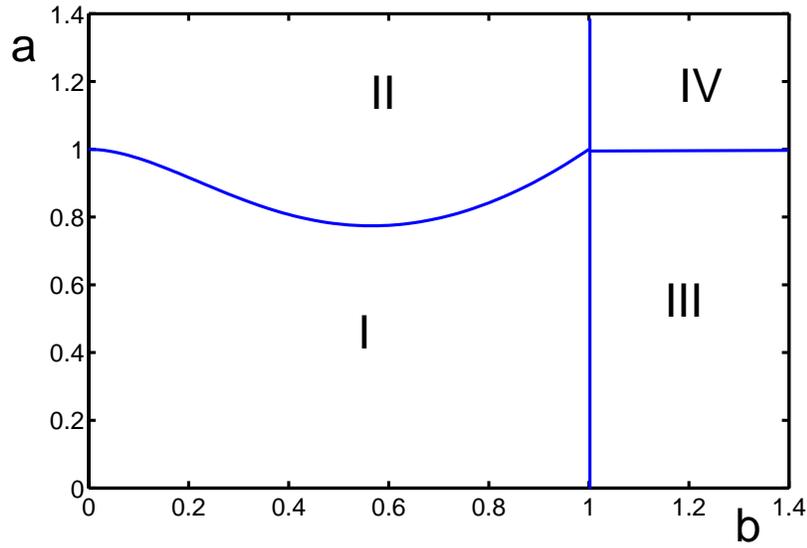} }
\caption{Phase portrait for the evolutionary stable states in the case of three
coexisting trait groups, cooperators, defectors, and regulators on the plane
of the relative amount of resources consumed by defectors $(a)$ and regulators
$(b)$. In region I, all three groups coexist, having nonzero fractions. In
region II, there exist only cooperators and regulators, while there are no
defectors $(x_2^* = 0)$. In region III, cooperators and defectors coexist,
but there are no regulators $(x_3^* = 0)$. In region IV, only cooperators are
present, without either defectors or regulators $(x_2^* = 0, x_3^* = 0)$.
}
\label{fig:Fig.2}
\end{figure}

\newpage

%Figura 3
\begin{figure}[ht]
\centerline{\includegraphics[width=10cm]{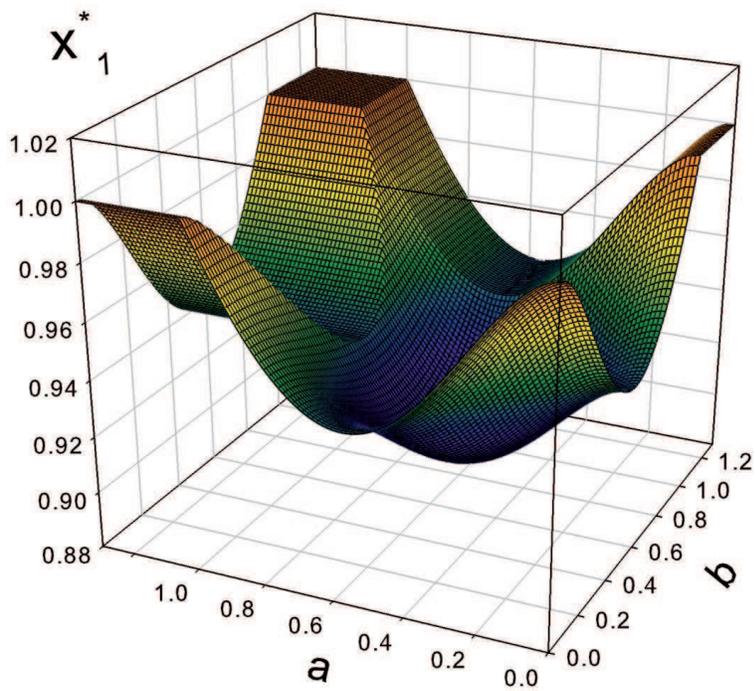} }
\caption{Evolutionary stable fraction of cooperators $x_1^*$ as a function
of the parameters $a$ and $b$.
}
\label{fig:Fig.3}
\end{figure}

\newpage

%Figura 4
\begin{figure}[ht]
\centerline{\includegraphics[width=10cm]{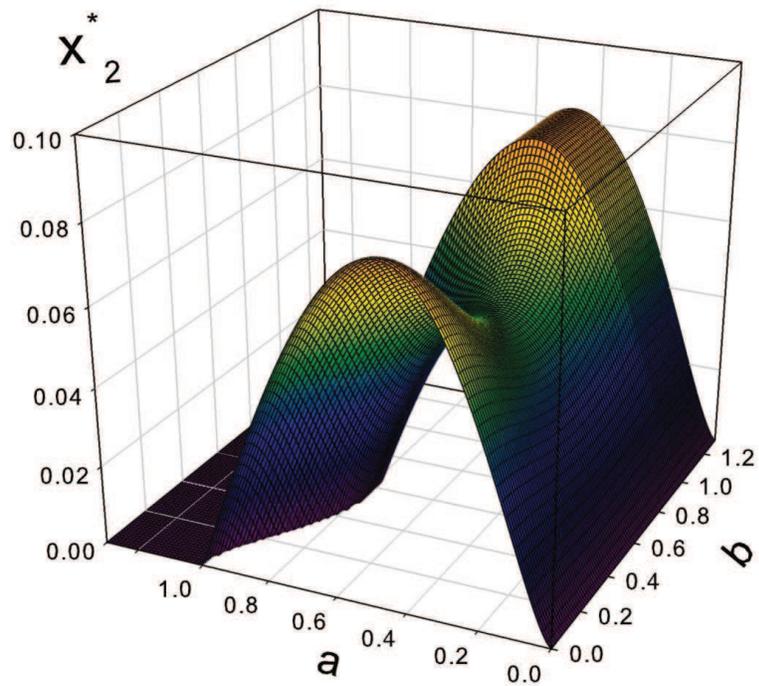} }
\caption{Evolutionary stable fraction of defectors $x_2^*$ as a function of
the parameters $a$ and $b$.}
\label{fig:Fig.4}
\end{figure}

\newpage

%Figura 5
\begin{figure}[ht]
\centerline{\includegraphics[width=10cm]{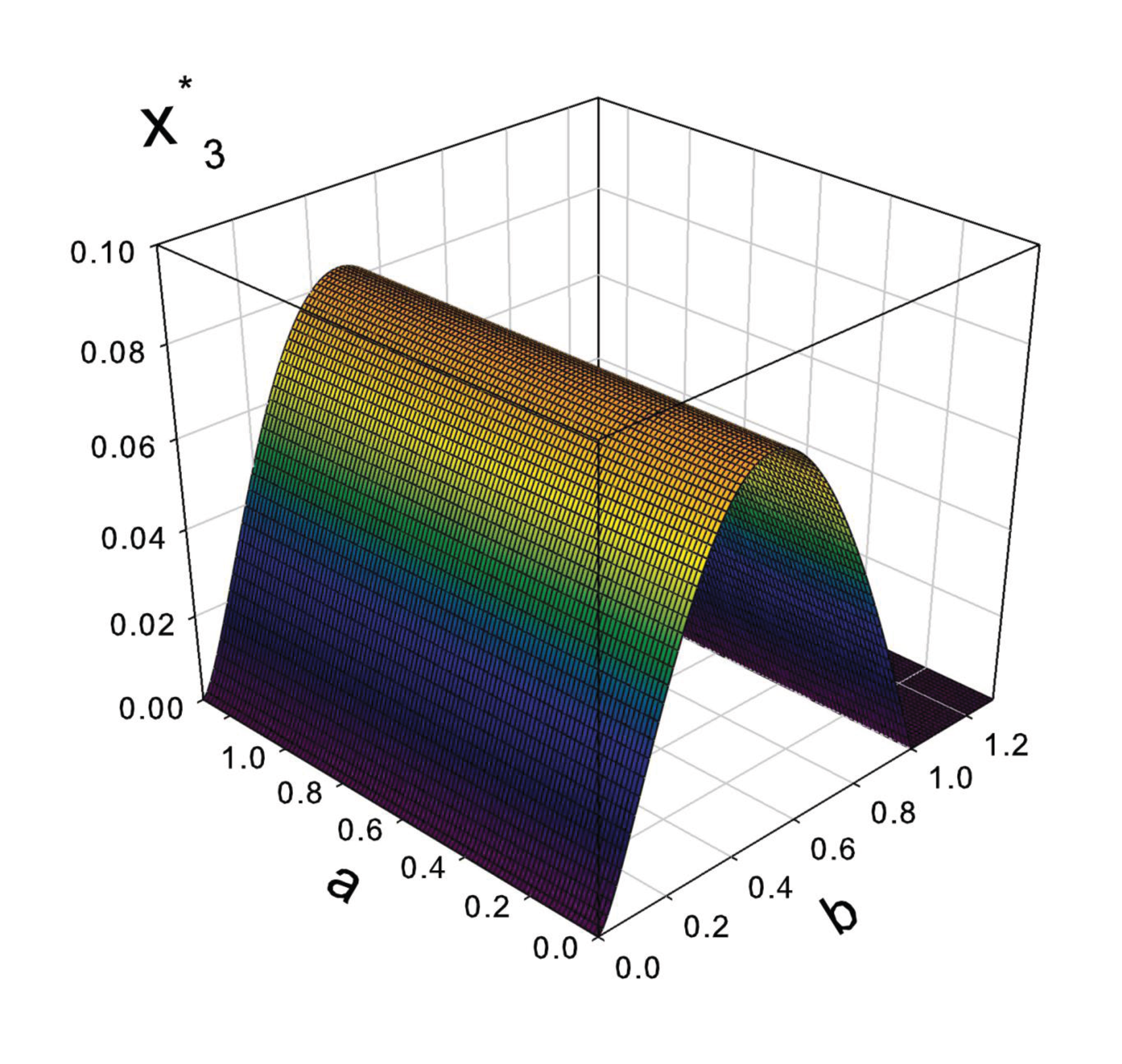} }
\caption{Evolutionary stable fraction of regulators $x_3^*$ as a function of
the parameters $a$ and $b$.}
\label{fig:Fig.5}
\end{figure}

\newpage

%Figura 6
\begin{figure}[ht]
\centerline{\includegraphics[width=15cm]{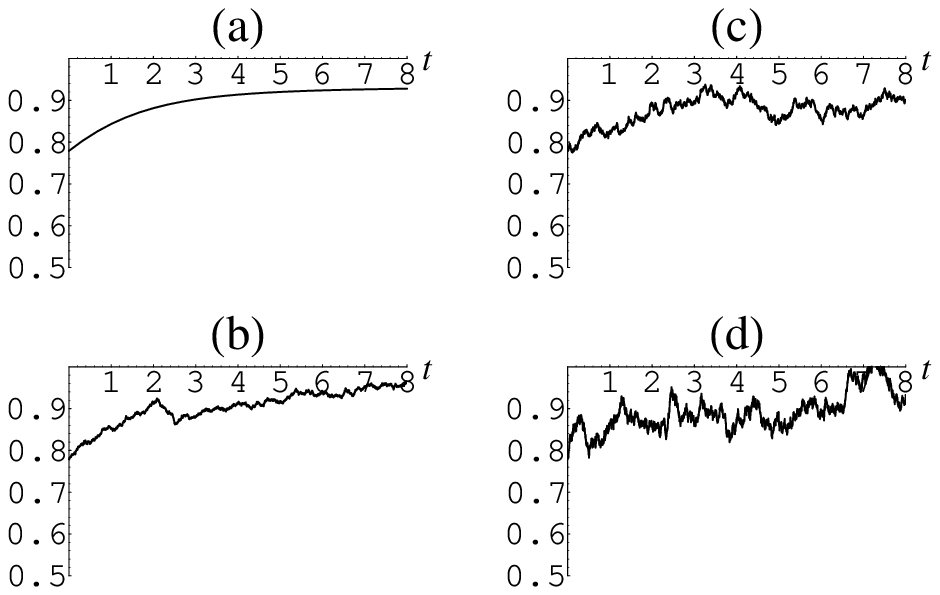} }
\caption{Temporal behavior of the cooperator fraction $x_1$ in the presence
of noise of different intensity for the parameters $a = 0.26$, $b = 0.33$
corresponding to Israel: (a) $\sigma = 0$; (b) $\sigma = 0.03$; (c) $\sigma = 0.06$;
(d) $\sigma = 0.09$.
}
\label{fig:Fig.6}
\end{figure}

\newpage

%Figura 7
\begin{figure}[ht]
\centerline{\includegraphics[width=15cm]{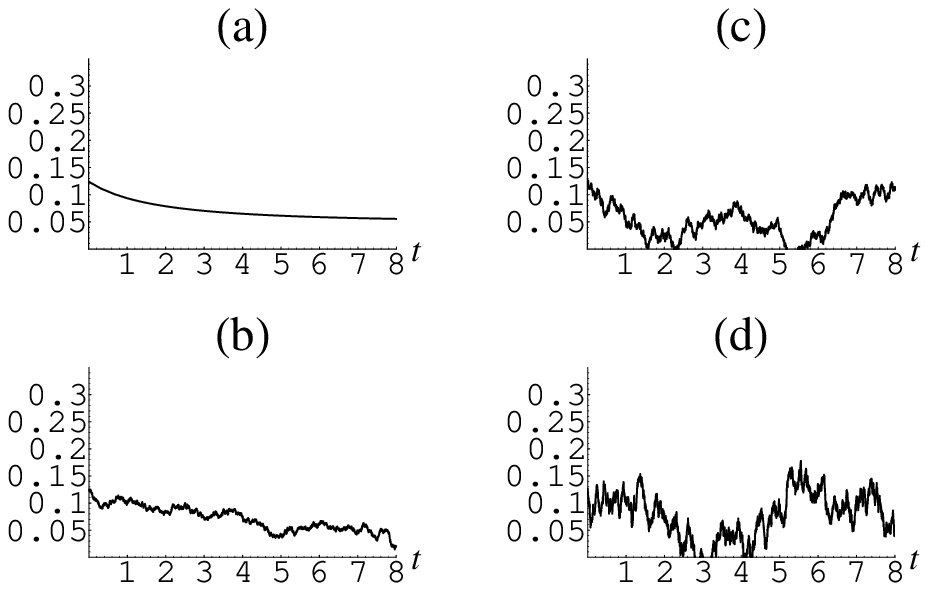} }
\caption{Temporal behavior of the defector fraction $x_2$ in the presence
of noise of different intensity for the parameters $a = 0.26$, $b = 0.33$
corresponding to Israel: (a) $\sigma = 0$; (b) $\sigma = 0.03$; (c) $\sigma = 0.06$;
(d) $\sigma = 0.09$.}
\label{fig:Fig.7}
\end{figure}

\newpage

%Figura 8
\begin{figure}[ht]
\centerline{\includegraphics[width=15cm]{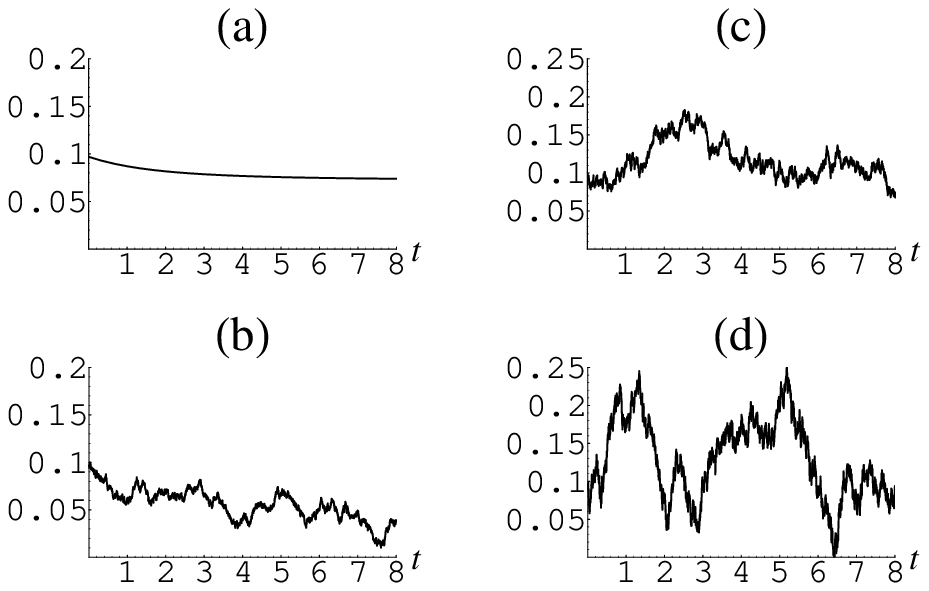} }
\caption{Temporal behavior of the regulator fraction $x_3$ in the presence
of noise of different intensity for the parameters $a = 0.26$, $b = 0.33$
corresponding to Israel: (a) $\sigma = 0$; (b) $\sigma = 0.03$; (c) $\sigma = 0.06$;
(d) $\sigma = 0.09$.} 
\label{fig:Fig.8}
\end{figure}

\end{document}